\documentclass[intlimits,twoside,a4paper]{article}

\usepackage{amsmath,amssymb}
\usepackage{graphicx}

\usepackage[T2A]{fontenc}
\usepackage[cp1251]{inputenc}

\usepackage[eqsecnum]{cmpj2}

\usepackage{color}

\issue{2013}{16}{3}{31702}
\doinumber{10.5488/CMP.16.31702}

\articletype{Proceedings Paper}

\title[Phase transition in (Ba$_{0.90}$Pb$_{0.10}$)(Ti$_{0.90}$Sn$_{0.10}$)O$_3$]
{Study of the phase transition in polycrystalline
(Ba$_{0.90}$Pb$_{0.10}$)(Ti$_{0.90}$Sn$_{0.10}$)O$_3$}

\author[C. Kajtoch, W. B\k{a}k, B. Garbarz-Glos]
{C. Kajtoch\thanks{E-mail:
ckajtoch@up.krakow.pl}\ , W. B\k{a}k, B. Garbarz-Glos}
\address{Institute of Physics, Pedagogical University, 2 Podchor\k{a}\.zych St., 30--084 Krak\'ow, Poland}

\date{Received  October 23, 2012, in final form November 9, 2012}
\authorcopyright{C. Kajtoch, W. B\k{a}k, B. Garbarz--Glos, 2013}

\begin{document}

\maketitle

\begin{abstract}
Polycrystalline sample of
(Ba$_{0.90}$Pb$_{0.10}$)(Ti$_{0.90}$Sn$_{0.10}$)O$_3$ was obtained by
means of a conventional ceramic technology. The dielectric
measurements were performed depending on temperature and frequency
of electric measuring field. The character of the phase
transitions of (Ba$_{0.90}$Pb$_{0.10}$)(Ti$_{0.90}$Sn$_{0.10}$)O$_3$
ceramics strongly depends on the presence of Pb in the sample.
The obtained results pointed out the diffused character of phase
transition. The temperature dependence of the dielectric
properties showed that the phase transition from the paraelectric
phase to ferroelectric one takes place at the same temperature
($T_\mathrm{m}=367$~K). It does not depend on the frequency of the
measuring electric field. A change of the value of the parameter
$\gamma$ takes place in the paraelectric phase.

\pacs 77.84.-s, 81.05.Je, 77.80.B, 77.90.+k
\end{abstract}

\section{Introduction}

Among a number of well-known ferroelectric materials,  barium
titanate BT (ABO$_3$-type compounds with perovskite structure)
and some of its solid solutions are the most interesting due to their
excellent dielectric properties. BaTiO$_3$ in its pure form does
not have ideal properties for industrial applications. The one
that has such properties is a barium lead stannate titanate solid
solution. The Pb-substitution at Ba-site is an effective way to
improve dielectric properties. A great attention is focussed on
Ba(Ti$_{1-x}$Sn$_x$)O$_3$ and (Ba$_{1-x}$Pb$_x$)TiO$_3$ solid solutions,
which are the most useful material for many applications and have
been extensively investigated, particularly their phase
transitions \cite{Kaj90,Mue04,Cro94,Kaj09,Kaj97,Kaj99,Sum09,Xin03}. It was
found that barium stannate titanate ceramics Ba(Ti$_{1-x}$Sn$_x$)O$_3$
exhibits many exceptional material properties and has a large range
of applications as a ceramic capacitor, PTCR thermistor,
piezoelectric transducers and actuators \cite{Hey61,Rav01}. The
paper presents the results of measurements of dielectric properties of
(Ba$_{0.90}$Pb$_{0.10}$)(Ti$_{0.90}$Sn$_{0.10}$)O$_3$ ceramics.

\section{Experimental}

The (Ba$_{0.90}$Pb$_{0.10}$)(Ti$_{0.90}$Sn$_{0.10}$)O$_3$ (abbreviated to
BP10TS10) polycrystalline sample was prepared by a solid state
synthesis. The sample was synthesized from analytically pure:
BaC$_2$O$_4$, PbC$_2$O$_4$, TiO$_2$ and SnO$_2$. The raw materials in
an appropriate molar ratio were ground and mixed in ether, then
dried and cold pressed. After the calcination at the
temperature of 1250~K and after re-milling, the BP10TS10
sample was sintered at the temperature of 1600~K for 2 hours.
A sample in the shape of disc-pellets sized 10~mm (diameter) and
1.50~mm thick was painted with silver electrodes.

The measurements were performed automatically using a LCR
Agilent 4284A meter and a temperature control system Quatro Krio
4.0 with and BDS 1100 cryostat. The research was done in the
frequency range from 20~Hz to 1~MHz under cooling process at rate
2~K/min.

\section{Results and discussion}

\begin{figure}[!b]
\centerline{
\includegraphics[width=0.45\textwidth]{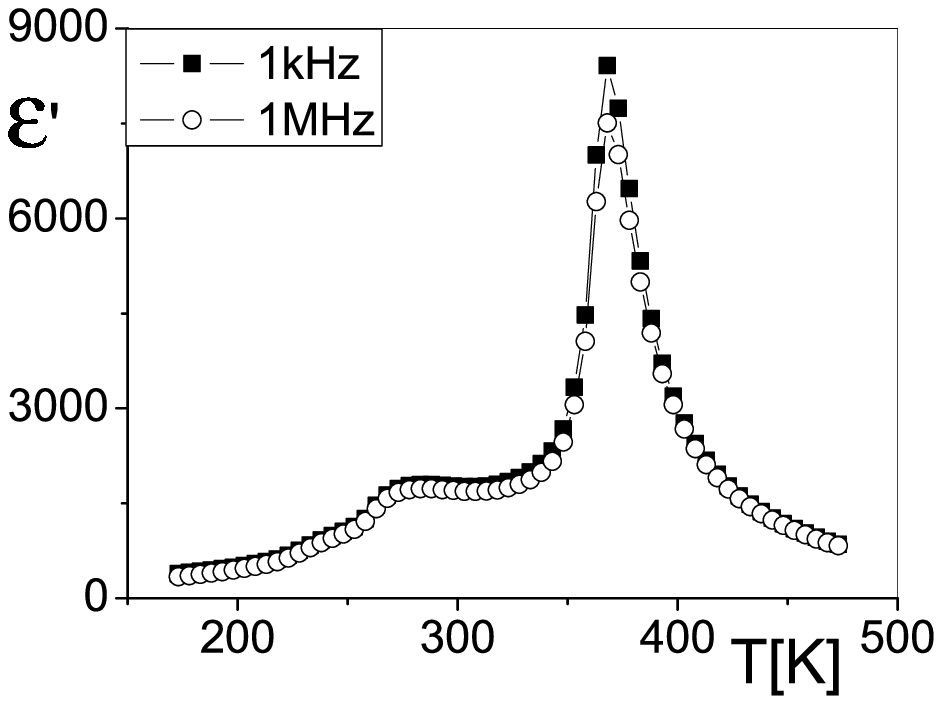}
\hspace{5mm}
\includegraphics[width=0.45\textwidth]{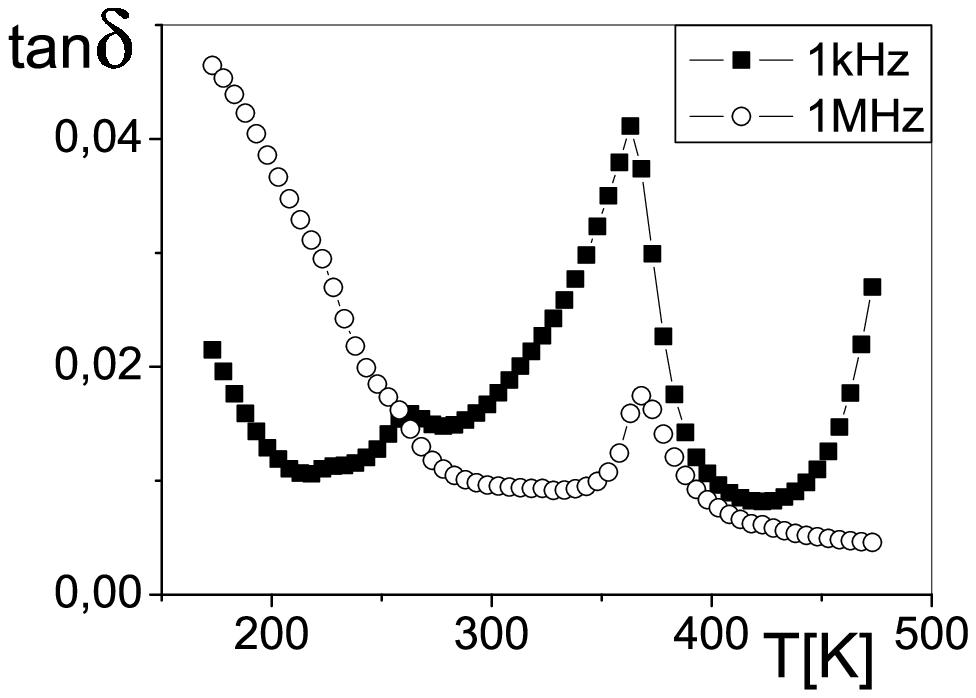}
}
\parbox[t]{0.48\textwidth}{%
\caption{ The dependence of the real part of electric permittivity
$\varepsilon'$ on temperature for the BP10TS10 sample.}
\label{fig1}
}%
\parbox[t]{0.48\textwidth}{%
\caption{ The dependence of the dielectric loss ($\tan \delta$) on
temperature for the BP10TS10.}
\label{fig2}
}
\end{figure}

The real part of the electric permittivity dependence on
temperature  $\varepsilon'(T)$ for BP10TS10 sample is presented
in figure~\ref{fig1}. For all frequencies of the electric field, the maximum
value of the electric permittivity  $\varepsilon'$ decreased, and
the temperature of the maximum ($T_\mathrm{m}$) did not change $T_\mathrm{m} =
367$~K.

The broadening of the temperature range of the phase transition
was also observed, which is connected with the degree of freezing
clusters during the transition from paraelectric to
ferroelectric phase in the cooling process. This behavior
indicates a diffusive nature of the phase transition.

Figure~\ref{fig2} shows the temperature dependence of dielectric loss
$\tan \delta$ during the cooling process. The dielectric loss
tangent exhibits a local anomaly in the vicinity of the
temperature of 367~K, which corresponds to the temperature
$T_\mathrm{m}$. The analysis of figure 2 shows that the local maxima
appear for all frequencies tested.

The effective polarization changes, thermally induced and
described as the  $\varepsilon'$ fast increase can be also
described by means of an electric modulus $M'$ ($M^* = M' + M''$)
dependence on temperature $T$. The modulus $M'$ is sensitive to
the small changes of local polarization.

The minima position in the $M'(T)$ curve can be interpreted as
temperature points representing transitions from one phase to another.

\begin{figure}[!t]
\centerline{
\includegraphics[width=0.45\textwidth]{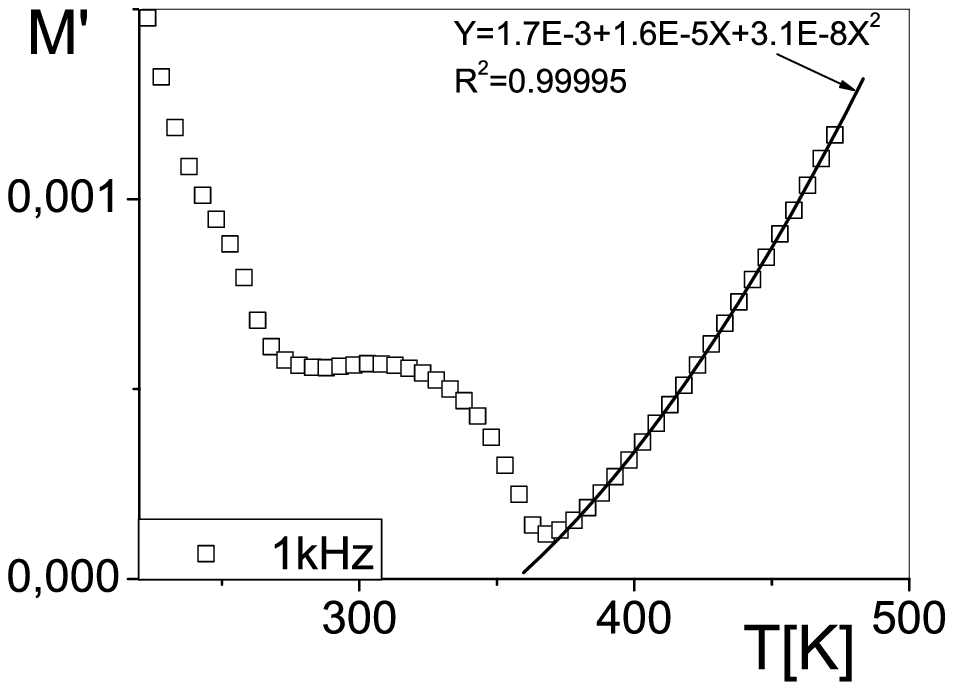}
\hspace{5mm}
\includegraphics[width=0.45\textwidth]{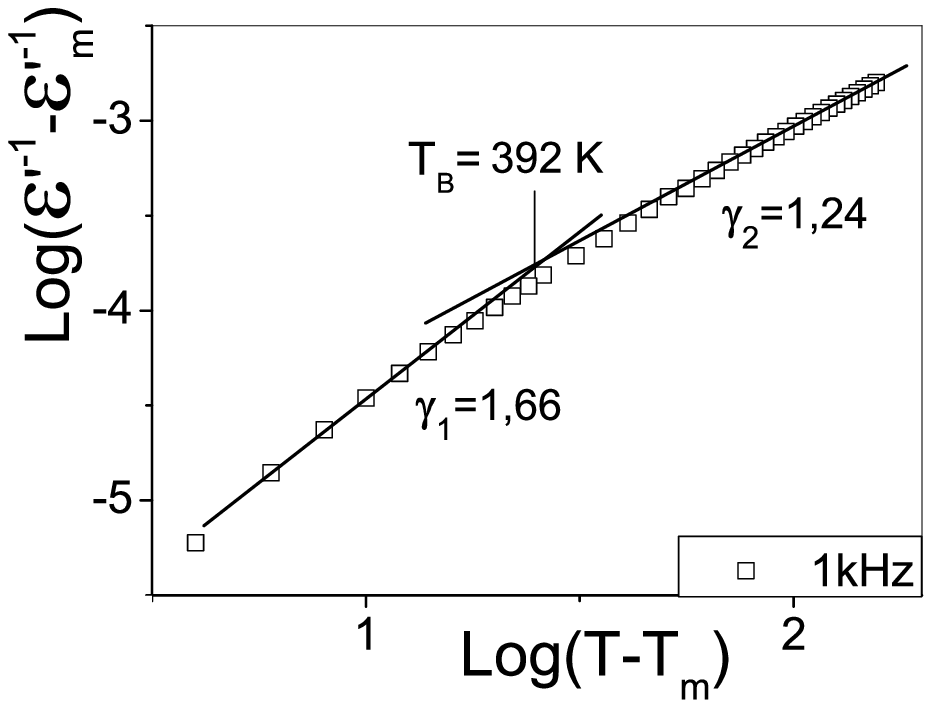}
}
\parbox[t]{0.48\textwidth}{
\caption{The dependence of the real part of electric modulus
($M'$) on temperature for the BP10TS10 sample.}
\label{fig3}
}
\parbox[t]{0.48\textwidth}{
\caption{ The dependence of
$\log(\varepsilon^{-1}-\varepsilon^{-1}_\mathrm{m})$ on $\log(T-T_\mathrm{m})$  for
the BP10TS10 sample.}
\label{fig4}

}

\end{figure}

Figure~\ref{fig3} presents the temperature dependence of the real part
of electric modulus ($M'$). In the paraelectric phase, the
nonlinear dependence was observed, which indicates a diffusive
nature of paraelectric-ferroelectric (PE-FE) phase transition.

One of the special properties of ferroelectric ceramics is
the diffuseness of the phase transition (diffuse phase transition~--- DPT). It means that the phase transition does not take place in
the whole specimen volume at a strictly determined Curie
temperature $T_\mathrm{C}$ (point phase transition), but in a certain
temperature zone (the so-called the Curie zone). Two structural phases,
one with lower and one with higher symmetry, coexist in this zone,
namely ferroelectric (low temperature phase) and paraelectric
(high temperature phase). The phenomenon of the diffuseness of the
ferroelectric phase transition has been discovered in both ceramic
materials and crystals and a lot of papers are devoted to
this phenomenon including the physical nature, causes, criteria
and the evaluation of the degree of diffuseness \cite{Smo70,Set80,Set82,Uch82,Isu89,San01}.

The following formula describes ferroelectric materials with
diffusive phase transition (DPT):
\begin{equation}
\label{eps} \varepsilon^{-1} = \varepsilon^{-1}_\mathrm{m} + A
(T-T_\mathrm{m})^{\gamma} \, ,\nonumber
\end{equation}
where:  $\varepsilon_\mathrm{m}$ is the maximum value of electric
permittivity; $T_\mathrm{m}$ is the temperature value at $\varepsilon_\mathrm{m}$; $A$
and $\gamma$ are constants for the chosen frequency. In DPT, the
value of $\gamma$ is close to 2 while for a sharp transition,
this value is close to 1. The values that follow from the above
formula are presented in figure~\ref{fig4} as a dependence of
$\log(\varepsilon^{-1}-\varepsilon^{-1}_\mathrm{m})$ on $\log(T-T_\mathrm{m})$.

The analysis of the obtained results indicates two temperature
regions with values $\gamma_1$ and $\gamma_2$. The values of these
parameters are 1.66 and 1.24, respectively. The change of the
value of $\gamma$ takes place in the region of the Burns
temperature $T_\mathrm{B} = 392$~K. This temperature is about 25~K
higher than the temperature of phase transition $T_\mathrm{m}$. The value
of $\gamma_1$ is close to 2 and suggests the behaviour of
ferroelectrics with DPT. The value of $\gamma_2$ is close to unity
and indicates a typical behavior for ferroelectrics with a sharp
phase transition at temperatures $T>T_\mathrm{B}$.

\begin{figure}[!b]
\centerline{
\includegraphics[width=0.45\textwidth]{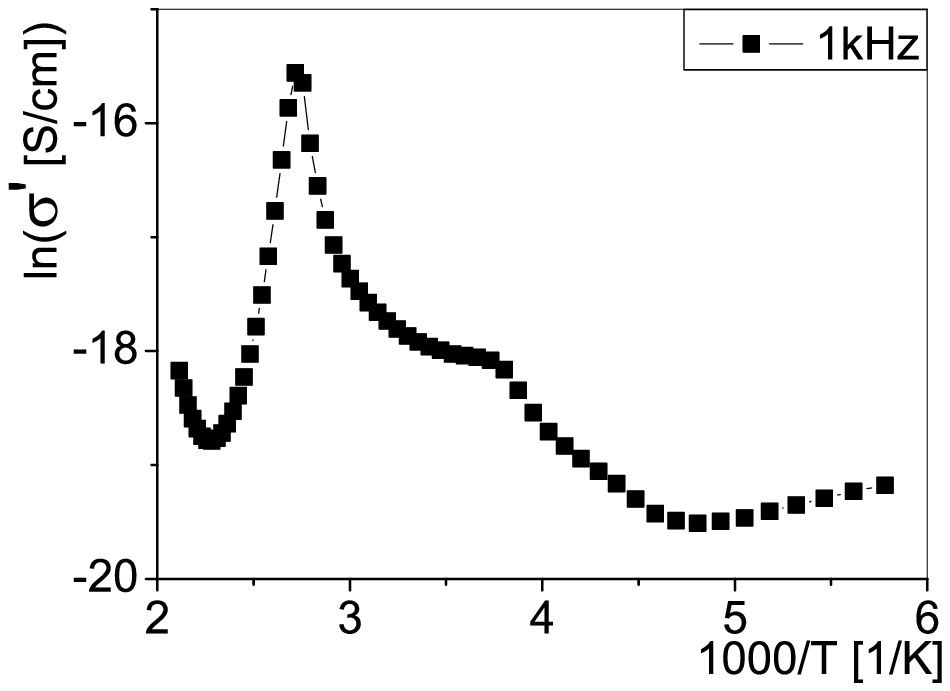}
\hspace{5mm}
\includegraphics[width=0.45\textwidth]{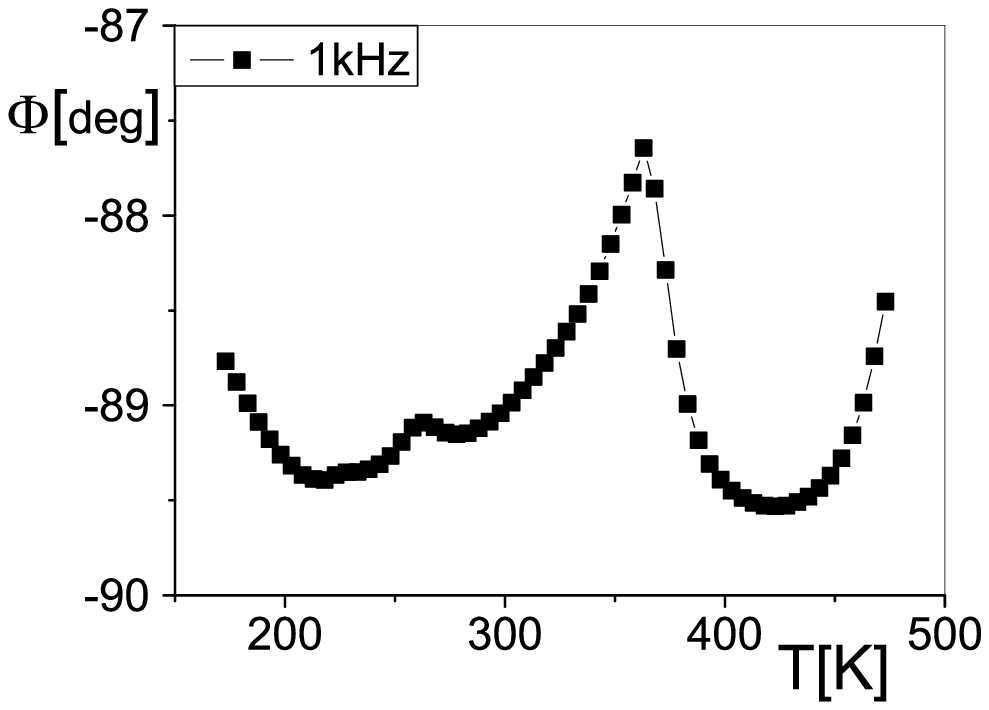}
}
\parbox[t]{0.48\textwidth}{
\caption{The ac-conductivity ($\sigma'$) dependence on reciprocal
temperature ($1000/T$) for the BP10TS10 sample.}
\label{fig5}
}
\parbox[t]{0.48\textwidth}{
\caption{The dependence of phase angle ($\phi$) on temperature for
the BP10TS10 sample.}
\label{fig6}
}
\end{figure}

Figure~\ref{fig5} presents the temperature changes of the real part of
electric conductivity $\sigma'$. The $(\ln \sigma')(1/T)$ curve
shows that the local maximum of a.c. conductivity occurs at the
temperature $T_\mathrm{m}$. Moreover, in the phase transition region, a
PTCR effect is observed.

The low values of phase angle $\phi$ in the investigated
temperature range (figure~\ref{fig6}) suggest the existence of polar
regions \cite{Kaj11, Ich72}, which contribute to the dipolar
polarization \cite{Bur82}.

The changes of cluster configuration leads to their liability and
sensitivity to the applied electric field. An increase of the value
of the phase angle at higher temperatures testifies to freeing
the charges \cite{Kaj99,Rav01,Ich72} and leads to an increase of
electric conductivity.

\section{Conclusions}

The dielectric studies show that the value of electric
permittivity of (Ba$_{0.90}$Pb$_{0.10}$)(Ti$_{0.90}$Sn$_{0.10}$)O$_3$
decreases with an increase of frequency in the whole investigated
temperature range.  The maximum of electric permittivity
$\varepsilon'$ is observed in the cooling process at a temperature of
367~K ($T_\mathrm{m}$). It was confirmed that the $T_\mathrm{m}$ does not depend
on the frequency of the measuring electric field. The obtained
dielectric data suggest a diffuse character of the phase
transition. The substitution of lead for barium in the amount of
10\% in the BP10TS10 ceramics compensates the effect of tin
on the temperature of the PE-FE phase transition and provides
high values of electric permittivity  $\varepsilon'$. The
occurrence of a Pb positional fluctuation in the paraelectric
phase is typically considered as a formation of polar regions. The
substitution of Pb for Ba causes a local distortion of the
lattice structure, resulting in a change of the electrostatic
forces (the long and short range). The change of the value of the
parameter  $\gamma$ in the paraelectric phase suggests a typical
behavior for  ferroelectrics with a sharp phase transition at
temperatures $T>T_\mathrm{B}$. The effect of differences in the values
of ionic radii and the deformation of a unit cell are reduced at
these temperatures.

Due to the existence of a PTCR effect in this compound, it can be
used as a material for thermistors. The obtained material is
expected to be a promising candidate for electronic ceramics.

\newpage

\ukrainianpart

\title{Вивчення фазового переходу в полікристалітах
(Ba$_{0.90}$Pb$_{0.10}$)(Ti$_{0.90}$Sn$_{0.10}$)O$_3$}

\author{Ц. Кайтох, В. Бак, Б. Гарбаж-Гльос}
\address{Інститут фізики, Педагогічний університет, Краків, Польща}

\makeukrtitle
\begin{abstract}

Полікристалічний зразок
(Ba$_{0.90}$Pb$_{0.10}$)(Ti$_{0.90}$Sn$_{0.10}$)O$_3$ отриманий за
допомогою стандартної керамічної технології. Діелектричні
вимірювання здійснені в залежності від температури і частоти
електричного вимірюючого поля. Характер фазових переходів керамік
(Ba$_{0.90}$Pb$_{0.10}$)(Ti$_{0.90}$Sn$_{0.10}$)O$_3$ сильно
залежить від присутності Pb в зразку. Отримані результати вказують
на дифузійний характер фазового переходу. Температурна залежність
діелектричних властивостей показала, що фазовий перехід з
параелектричної  до сегнетоелектричної фази має місце при тій самій
температурі ($T_\mathrm{m}=367$~K). Вона не залежить від частоти
вимірюючого електричного поля.  Зміна значення параметра $\gamma$
має місце в параелектричній фазі.

\keywords діелектричні властивості, фазовий перехід, полярні
області, кераміка
\end{abstract}\end{document}